\documentclass[amsmath,amssymb,aps,prl,reprint,superscriptaddress]{revtex4-1}

\usepackage{graphicx}

\let\vaccent=\v 
\renewcommand{\v}[1]{\ensuremath{\mathbf{#1}}} 

\newcommand{\abs}[1]{\left| #1 \right|} 

\newcommand{\ket}[1]{\left| #1 \right>} 
\newcommand{\bra}[1]{\left< #1 \right|} 
\newcommand{\matrixel}[3]{\left< #1 \vphantom{#2#3} \right|#2 \left| #3 \vphantom{#1#2} \right>} 
\DeclareMathOperator{\Tr}{Tr}

\begin{document}


\title{Quantum-polarization state tomography}


\author{\"Omer Bayraktar}
\altaffiliation{Present address: Max Planck Institute for the Science of Light, 91058 Erlangen, Germany}
\email[]{oemer.bayraktar@mpl.mpg.de}
\affiliation{Department of Applied Physics, Royal Institute of Technology  (KTH), 10691 Stockholm, Sweden}
\affiliation{Physik-Department, Technische Universit\"at M\"unchen, 85748 Garching bei M\"unchen, Germany}

\author{Marcin Swillo}
\affiliation{Department of Applied Physics, Royal Institute of Technology (KTH), 10691 Stockholm, Sweden}

\author{Carlota Canalias}
\affiliation{Department of Applied Physics, Royal Institute of Technology (KTH), 10691 Stockholm, Sweden}

\author{Gunnar Bj\"ork}
\affiliation{Department of Applied Physics, Royal Institute of Technology (KTH), 10691 Stockholm, Sweden}


\date{\today}

\begin{abstract}
We propose and demonstrate a method for quantum-state tomography of qudits encoded in the quantum polarization of $N$-photon states. This is achieved by distributing $N$ photons nondeterministically into three paths and their subsequent projection, which for $N=1$ is equivalent to measuring the Stokes (or Pauli) operators. The statistics of the recorded $N$-fold coincidences determines the unknown $N$-photon polarization state uniquely. The proposed, fixed setup manifestly rules out any systematic measurement errors due to moving components and allows for simple switching between tomography of different states, which makes it ideal for adaptive tomography schemes.
\end{abstract}


\maketitle

Quantum-state tomography is related to the Pauli problem, i.e., to determining quantum states from measurements~\cite{Weigert1992}. The initial ideas for a solution of the Pauli problem were stated by Fano in 1957, who coined the term \emph{quorum}, which denotes a set of observables whose measurements provide tomographically complete information about a system \cite{Fano1957}. Measuring a quorum on a finite ensemble of a system, the quantum state of the system is inferred. The first method for quantum-state tomography was developed for continuous variables~\cite{Vogel1989}. The question whether the quantum-state tomography of discrete systems is also possible \cite{Leonhardt1995} was answered by experimentally measuring the quantum analog of Stokes parameters \cite{Stokes1856} from which the quantum-polarization state of identically prepared photonic qubits were inferred \cite{White1999}. Subsequently, the theory of qubit \cite{James2001} and qudit \cite{Thew2002} quantum state tomography was studied. Today, quantum-state tomography is an indispensable benchmark in experimental quantum information with continuous variables and qubits~\cite{Lvovsky2009,Versteegh2015,Lee2015a,McKay2015}. Concurrently, improved security in quantum cryptography \cite{Scarani2009}, computational speed-up \cite{Gedik2015}, increased resolution in quantum metrology \cite{Bjork2015a}, and more fundamental questions \cite{Brunner2014} have increasingly drawn attention to qudits. However, implementations of qudit-state tomography have been only recently applied to physical systems such as nuclear spins \cite{Miranowicz2015}, the orbital angular momentum degree of freedom of photons \cite{Giovannini2013, Bent2015}, and low-dimensional optical qudits \cite{Bogdanov2004, Kwon2014}. A general procedure for the quantum-state tomography of optical qudits in arbitrary dimensions implemented by polarization states, despite their significance, has yet to be proposed. In this Rapid Communication we present such an experimental procedure. The result is a compact, self-calibrating setup without moving components. Consequently, the influence of systematic measurement errors is reduced to a minimum. The results presented here constitute a generalization of previous approaches in quantum-polarization state tomography \cite{White1999, Bogdanov2004} and provide a benchmarking tool for experiments exploiting the quantum polarization of multiphoton states.

\textit{Theory.}
The quantum state of a $d$-dimensional system is represented by a qudit density matrix $\hat\rho$ which can be linearly expanded in terms of a set of $d^2-1$ operators $\hat\lambda_\nu$ as
\begin{equation}
\hat\rho= \frac{\hat{1}}{d} + \sum_{\nu=1}^{d^2-1}\lambda_\nu \hat\lambda_\nu \;.
\label{Eq: Decomposition}
\end{equation}
Here, the symbol $\hat 1$, in the following replaced by $\hat\lambda_0$, represents the identity operator in $d$ dimensions and the operators $\hat\lambda_\nu$ are traceless generators of the SU($d$) group. A common choice is to pick the generalized Pauli operators \cite{Thew2002}. When projecting the state onto a measurement projector $\ket{\psi_\mu}\bra{\psi_\mu}$, it follows trivially that the probability of finding the system $\hat\rho$ in the state $\ket{\psi_\mu}$ $(\mu=1,\dotsc,d^2-1)$ after a measurement is given by
\begin{equation}
P_{\ket{\psi_\mu}}=\matrixel{\psi_\mu}{\hat\rho}{\psi_\mu}\;.
\end{equation}
Measuring a large number of ensemble members of $\hat\rho$, the probabilities $P_{\ket{\psi_\mu}}$ can be estimated. Then, a linear system of equations relating the probabilities $P_{\ket{\psi_\mu}}$ with the generators can be constructed \cite{Thew2002},
\begin{equation}
P_{\ket{\psi_\mu}}= \sum_{\nu=0}^{d^2-1}B_{\mu\nu}\lambda_\nu\;,
\label{Eq: Probabilities}
\end{equation}
and, under the condition that the matrix $B_{\mu\nu}\equiv\matrixel{\psi_\mu}{\hat\lambda_\nu}{\psi_\mu}$ is nonsingular, Eq.~\eqref{Eq: Probabilities} can be inverted in order to determine the expansion coefficients $\lambda_\nu$, which in turn can be inserted into Eq.~\eqref{Eq: Decomposition} to obtain the state's density operator. In this case the set of projectors $\ket{\psi_\mu}\bra{\psi_\mu}$ is complete. Once such a set is determined for a single-qudit state, a generalization towards multiple-qudit states is straightforward: Applying local measurement projections of each single qudit into a complete set of states and measuring all single and joint probabilities between local measurements of different qudits, a tomographic reconstruction of the state is possible \cite{Wootters1990}. However, in prime dimensional Hilbert spaces the density matrix cannot be factored into subsystems such as qubits or qutrits, and even if the dimensionality of the system allows factorization, it is often difficult to generate arbitrary states by using a tensor space of qubits and qutrits, as most states will then be entangled.

We initially treat the case that the qudits are physically implemented by $N$-photon polarization states. (The tomography of polarization states with an indeterminate number of photons will be discussed later.) A pure state of that kind can be written
\begin{equation}
\ket\psi=\sum_{m=0}^N{\alpha_m\ket{N-m,m}}\;,
\end{equation}
where $\ket{N-m,m}$ stands for $N-m$ photons in the horizontal polarization mode and $m$ photons in the vertical polarization mode, and the complex coefficients $\alpha_m$ fulfill $\sum\nolimits_{m=0}^N \abs{\alpha_m}^2=1$. This state is equivalent to a single qudit with $d=N+1$ dimensions.

To the best of our knowledge, tomography strategies of optical qudits have been developed and implemented successfully for photon numbers $N=1$ \cite{White1999} and $N=2$ \cite{Bogdanov2004} only. Especially, the case of $N=1$, i.e., optical qubit-state tomography, is widely employed and an indispensable tool in experimental quantum information. Our tomography proposal extends the ability to tomograph to, in principle, any two-mode qudit system by using physically local, but Hilbert-space nonlocal, measurements. Although in general such a strategy is difficult to implement, it turns out that a compact setup exists that is ideally suitable as a solution. Here we demonstrate it by using $N$-photon polarization states.

\begin{figure}
\includegraphics[width=1.00\columnwidth]{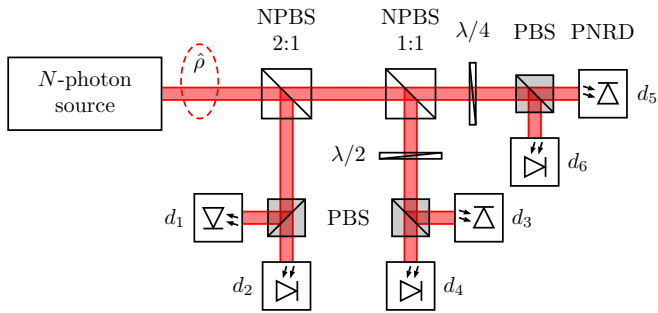}
\caption{\label{fig:setup} Schematic drawing of the experimental setup implementing our protocol for $N$-photon polarization state tomography. The incoming beam of $N$-photons is divided into three paths. This is accomplished by sending the $N$-photons through a nonpolarizing beam splitter reflecting 1/3 and transmitting 2/3 (NPBS 2:1) and their subsequent splitting into two other equal parts by a symmetric nonpolarizing beam splitter (NPBS 1:1). On the single-photon level the projective measurements in the distinct paths realized by combinations of wave retarders, polarizing beam splitters (PBS), and photon-number resolving detectors (PNRD), correspond to applying the Stokes operators $\hat S_x,\hat S_y$ and $\hat S_z$ \cite{James2001}, i.e., the half-wave ($\lambda/2$) and quarter-wave ($\lambda/4$) retarders are oriented at $\tfrac{\pi}{8}$ and $\tfrac{\pi}{4}$, respectively. Recording the statistics of $N$-fold coincident photon arrivals provides all necessary data for the reconstruction of an unknown $N$-photon polarization state $\hat\rho$.}
\end{figure}

An implementation of our proposed measurement setup is depicted in Fig.~\ref{fig:setup}.
It resembles the setup used in classical (first-order) polarimetry, i.e., in the determination of Stokes parameters or equivalently optical qubit-state tomography \cite{Stokes1856,James2001}. However, while in classical polarimetry analog detectors measure macroscopic light intensities, and in optical qubit-state tomography single-photon detectors are used merely for reasons of sensitivity, in our scheme we use photon-number resolving detectors in order to display all possible correlations. Such detectors are coming of age, and are becoming commercially available. Initially in the analysis we will assume that the detectors have unit quantum efficiency. Later we shall show that this assumption can be dropped.

The number of photons counted by each detector $i$ ($i=1,\dotsc,6$) is denoted by $d_i$ with $\sum_{i=1}^6 d_i=N$. The vector $(d_1,\dotsc,d_6)$ is called an event and corresponds to a projector. An $N$-photon state may give rise to
\begin{equation}
\mathcal{M}(N)=\left(
\begin{array}{c}
	N+5\\
	N
\end{array}
\right)
\end{equation}
different events, which can be resolved by $N$-fold coincidence detection. Note that $\mathcal{M}(N) \propto N^5$, so the number of projectors increases very dramatically with the space dimension $N$. The respective projectors corresponding to the events can be calculated by further developing the ideas presented in Refs.~\cite{Hofmann2004, Li2008}. Then, the projected state as a function of the number of photons arrived at the detectors $d_i$ reads
\begin{widetext}
\begin{equation}
\ket{\psi_\mu}=\frac{\left(\hat a_\mathrm{V}^\dagger\right)^{d_1}\left(\hat a_\mathrm{H}^\dagger\right)^{d_2}\left(\hat a_\mathrm{V}^\dagger+\hat a_\mathrm{H}^\dagger\right)^{d_3}\left(\hat a_\mathrm{V}^\dagger-\hat a_\mathrm{H}^\dagger\right)^{d_4}\left(i\hat a_\mathrm{V}^\dagger+\hat a_\mathrm{H}^\dagger\right)^{d_5}\left(i\hat a_\mathrm{V}^\dagger-\hat a_\mathrm{H}^\dagger\right)^{d_6}\ket{0}}{\sqrt{\mathcal{N}(d_1,d_2,d_3,d_4,d_5,d_6)}}\;,
\label{eq:projectingStates}
\end{equation}
\end{widetext}
where $\hat a^\dagger_\mathrm{V}$ and $\hat a^\dagger_\mathrm{H}$ are the creation operators for vertically and horizontally polarized photons, respectively, $\ket{0}$ is the vacuum state, and $\mu$ is shorthand for the event vector $\left(d_1,d_2,\dotsc,d_6\right)$. Calculation of the normalization factor requires special care due to the fact that, after separation into different paths and subsequent detection, photons become distinguishable. We find that $\mathcal{N}(\mu)=d_1!d_2!\mathcal{N}_2(d_3,d_4)\mathcal{N}_2(d_5,d_6)$ with
\begin{widetext}
\begin{equation}
\mathcal{N}_2(d_i,d_j)=\sum_{d_i',d_i''=0}^{d_i}\sum_{d_j',d_j''=0}^{d_j}
\left(\begin{array}{c}
	d_i\\
	d_i'
\end{array}\right)
\left(\begin{array}{c}
	d_i\\
	d_i''
\end{array}\right)
\left(\begin{array}{c}
	d_j\\
	d_j'
\end{array}\right)
\left(\begin{array}{c}
	d_j\\
	d_j''
\end{array}\right)
\left(-1\right)^{d_j'+d_j''}\left[d_i+d_j-\left(d_i'+d_j'\right)\right]!\left(d_i'+d_j'\right)!\delta_{d_i'+d_j',d_i''+d_j''}\;.
\end{equation}
\end{widetext}
The completeness of the projected states $\ket{\psi_\mu}$ with respect to the unique determination of the unknown state $\hat\rho$ follows from calculating the rank $r$ of the nonsquare matrix $B_{\mu\nu}$. We have found that the rank is $r=d^2=(N+1)^2$ for $N=2$ to 7, and we conjecture that this is the case for any $N \geq 2$. If so, two conclusions follow: The rank is sufficient to determine $\hat{\rho}$ by inverting the linearly independent part of $B_{\mu\nu}$. However, since $\mathcal{M}>(N+1)^2$, the number of projected states $\ket{\psi_\mu}$ is overcomplete, making the measurement data perfectly compatible with maximum-likelihood estimation methods (MAXLIK) \cite{Hradil1997}. Therefore, under the made assumption of unit quantum efficiency we conjecture that by using the experimental setup depicted in Fig.~\ref{fig:setup}, quantum-state tomography of qudits encoded in the polarization degree of freedom of an $N$-photon state can be performed for an arbitrary $N$. Then, also multiple-qudit states can be reconstructed based on the discussed experimental setup, e.g., multiple, spatially distinguishable $N$-photons are tomographed by applying the setup in Fig.~\ref{fig:setup} locally in each spatial mode.

In practice, detectors have nonunity quantum efficiency. Moreover, for photon-number resolving detectors it is a function of the number of detected photons, i.e., detector $i$'s efficiency for detecting $k$ photons of $n$ impinging photons is $\eta_i(k|n)$. For states with photon number $N$, by $N$-fold coincidence detection, only efficiencies $\eta_i(n|n)\equiv\eta_i(n)$ are adequate. However, by this strategy events with lost photons are ignored in the state estimation process, while including them will increase the number of possible events. Thus, in the quantum-state tomography of $N$-photon states, there will be $6 N$ additional independent unknowns due to nonunity quantum efficiencies. We have tabulated the number of unknown parameters and the number of projectors in Table~\ref{Table: Parameters}.
\begin{table}
\begin{ruledtabular}
\begin{tabular}{cccccc}
\textrm{$N$}&
\textrm{$\hat{\rho}$ parameters}&
\textrm{\begin{tabular}{@{}c@{}}Quantum\\ efficiencies\end{tabular}}&
\textrm{$\mathcal{M}(N)$}&
\textrm{\begin{tabular}{@{}c@{}}Unknowns\\ up to $N$\end{tabular}}&
\textrm{$\sum_0^N \mathcal{M}(N)$}\\
\colrule
  0 & 1 & 0 & 1 & 1 & 1 \\
  1 & 4 & 6 & 6 & 11 & 7 \\
  2 & 9 & 12 & 21 & 26 & 28 \\
  3 & 16 & 18 & 56 & 48 & 84 \\
  4 & 25 & 24 & 126 & 79 & 210 \\
  $n$ & $(n+1)^2$ & $6n$ & $\frac{(n+5)!}{n! 5!}$ & $O(n^3)$ & $O(n^6)$ \\
\end{tabular}
\end{ruledtabular}
\caption{\label{Table: Parameters} Count of parameters in the estimation process as a function of the polarization state's photon number $N$. The total number of unknowns up to excitation manifold $N$ in column 5 is given by summing the numbers in column 2 over the photon numbers $0,1,\dots, N$ and adding the number of unknown quantum efficiencies.}
\end{table}
From the table it is evident that for $N \geq 2$ the number of projectors (in column 4) rapidly becomes much larger than the number of unknowns, including the quantum efficiencies of the detectors (sum of column 2 and 3). Consequently, the quantum efficiencies can be included as parameters in the MAXLIK estimation, but in this case the authors could not guarantee that the reconstruction will give a unique result. The uniqueness of the reconstructed state is restored by precalibrating the quantum efficiencies. In addition, for $N \geq 2$, the total number of projectors up to a photon number $N$ (column 6) is larger than the total number of unknowns (column 5). Note that the projectors parse an initially unknown quantum state into polarization sectors with fixed photon numbers $N$. Consequently, the quantum state is accessed in its block-diagonal form only \cite{Bjork2015a}. Thus, the presented tomography scheme could allow for the state reconstruction with \textit{a priori} unknown $N$, and also mixtures of different photon-number states. In practice, however, our scheme will be limited to Hilbert space dimensions of the order ten, simply because of the staggering amount of data needed to provide a detailed density-matrix reconstruction of high-dimensional states.

\textit{Experimental methods.}
In order to demonstrate the feasibility of the proposed method, we have performed quantum-state reconstruction of optical qutrits. The two-photon source consists of a narrow-bandwidth continuous-wave diode laser with a central wavelength at 405 nm pumping a periodically poled Potassium Titanyl Phosphate nonlinear optical crystal. The poling period of 10.1 $\mu$m was chosen to allow for an efficient generation of spatially and spectrally degenerate signal and idler photon pairs at 810 nm in type-II spontaneous parametric down-conversion around room temperature. After the temporal walk-off between signal and idler photons due to the birefringence of the nonlinear optical crystal is compensated in a polarization Michelson interferometer, the quantum state of the photon-pairs behind a half-wave plate making an angle $\theta/2$ with the horizontal axis is \cite{Shih2003}
\begin{eqnarray}
\ket\psi=&&-\sqrt{2}\cos\theta\sin\theta\ket{2,0}\nonumber\\
&&+\cos\left(2\theta\right)\ket{1,1}+\sqrt{2}\cos\theta\sin\theta\ket{0,2}\;.
\end{eqnarray}

On the detection side we have combined "on-off" single-photon avalanche diodes in spatial multiplexing in order to render possible the detection of photon numbers up to two \cite{Paul1996}. Furthermore, instead of collecting all $\mathcal{M}$ possible events simultaneously, they were sampled consecutively using optical switching. Therefore, using only two off-the-shelf single-photon detectors, all required measurements for $N=2$ can be performed, underlining the fact that the presented tomographic method can be readily applied using minimal resources. The generated electronic pulses from the detectors, triggered by photon arrivals, are acquired by a time-to-digital converter. Then, in postprocessing, coincidences between the detectors for distinct optical-switch configurations are extracted and the number of counts $n_\mu$ resulting in event $\mu$ are obtained. As discussed above, non-unity detection efficiencies are introduced as parameters such that the expected number of counts from a state $\hat\rho$ reads
\begin{equation}
\bar n_\mu=I\prod_{i=1}^6 \eta_i(d_i) P_{\ket{\psi_\mu}}\;,
\end{equation}
with $I$ the measured ensemble size, and are determined in the maximum-likelihood estimation as well as the unknown state $\hat\rho$ through numerical minimization of the penalty function \cite{Altepeter2004}
\begin{equation}
\chi^2=\sum_{\mu}\frac{\left(\bar n_\mu-n_\mu\right)^2}{\bar n_\mu}\,.
\end{equation}
A constraint ensuring non-negativity of the state $\hat\rho$ is imposed through its representation in form of a Cholesky decomposition \cite{Altepeter2004}
\begin{equation}
\hat\rho=\frac{\hat T^\dagger\hat T}{\Tr\left[\hat T^\dagger\hat T\right]}\;,
\end{equation}
where $\hat T$ is a lower triangular matrix. The resulting state from the minimization process is the one with the highest probability to result in the measured counts $n_\mu$.

\begin{figure*}
\includegraphics[width=2.00\columnwidth]{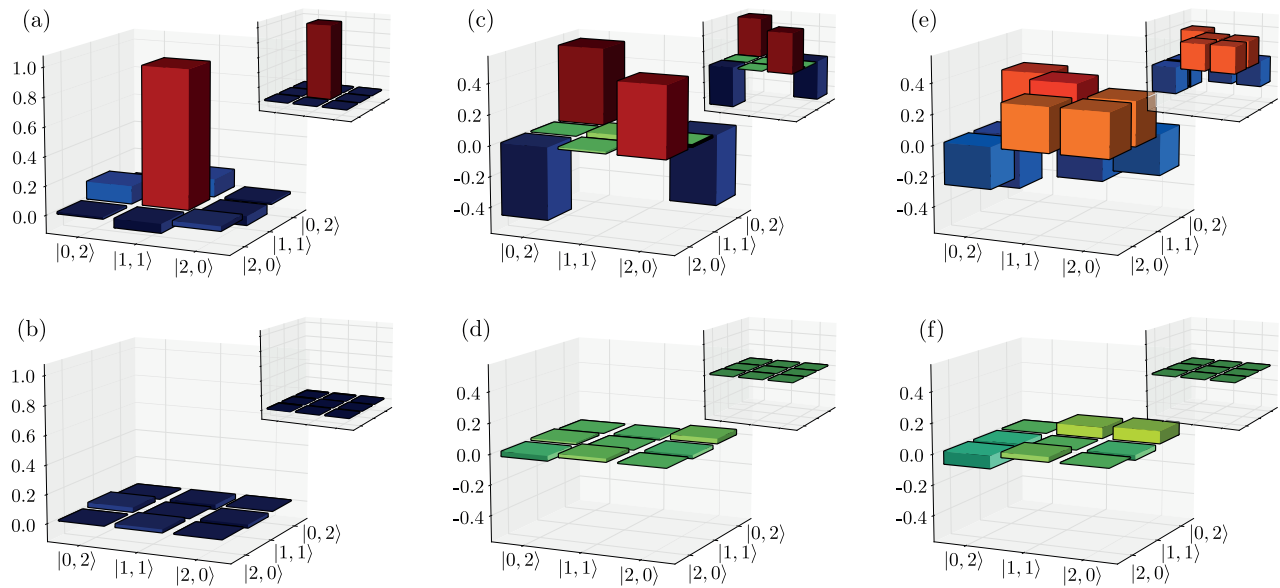}%
\caption{\label{fig:results} Graphical representation of the reconstructed states using the proposed tomography method. The (a) real and (b) imaginary parts of the entries of the reconstructed $\ket{1,1}$ state are shown. The insets display real and imaginary parts of the target states for reference. The fidelity between the target and the reconstructed state is $F=0.95\pm0.03$. In case of the two-photon NOON state [(c) and (d)], $F=0.960\pm0.004$. For the equipartition state $(-\ket{2,0}+\ket{1,1}+\ket{0,2})/\sqrt{3}$ [(e) and (f)], $F=0.932\pm0.003$.}
\end{figure*}

We have performed the described procedure of quantum-state tomography and reconstruction for several orientations of the half-wave retarder, i.e., for different prepared states. The measured ensemble size of each of the states was around 50\,000. For the half-wave retarder oriented at $\theta=0$, the prepared state is ideally $\ket{1,1}$ and the preparation and subsequent tomography process resulted in the estimation
\begin{equation}
\hat\rho=
\left(\begin{array}{ccc}
	0.02 & 0.12-0.03i & -0.01 \\
	0.12+0.03i & 0.96 & -0.06-0.02i \\
	-0.01 & -0.06+0.02i & 0.03
\end{array}\right)\;.
\end{equation}
In order to quantify the agreement between expectations and experimental results, we calculate the fidelity $F\equiv\matrixel{\psi}{\hat\rho}{\psi}$ of the target state $\ket\psi$ with ten generated and reconstructed states $\hat\rho$ obtained from performing the maximum-likelihood estimation on random variates of the measured counts according to Poissonian statistics, and give mean and standard deviation. Here, the fidelity is estimated to $F=0.95\pm0.03$. When repeating the measurements for the two-photon NOON state $(\ket{2,0}-\ket{0,2})/\sqrt{2}$ ($\theta=\pi/4$), we have obtained
\begin{equation}
\hat\rho=
\left(\begin{array}{ccc}
	0.51 & 0.00-0.01i & -0.47+0.03i \\
	0.00+0.01i & 0.03 & 0.01-0.02i \\
	-0.47-0.03i & 0.01+0.02i & 0.46
\end{array}\right)\;,
\end{equation}
and the fidelity $F=0.960\pm0.004$ with the state we had the intention to prepare. Orienting the half-wave retarder at $\theta=0.076\pi$, the two-photon polarization state $\ket{1,1}$ is in theory transformed into the perfect equipartition state $(-\ket{2,0}+\ket{1,1}+\ket{0,2})/\sqrt{3}$. In this case the reconstruction procedure gave
\begin{equation}
\hat\rho=
\left(\begin{array}{ccc}
	0.34 & -0.35+0.07i & -0.27+0.08i \\
	-0.35-0.07i & 0.37 & 0.29-0.03i \\
	-0.27-0.08i & 0.29+0.03i & 0.29
\end{array}\right)\;,
\end{equation}
with an estimated fidelity of $F=0.932\pm0.003$ with an ideal equipartition state.
Graphical representations of the reconstructed and prepared states are shown in Fig.~\ref{fig:results}.
The results display a good agreement with the prepared states. The nonunity fidelities are due to imperfections in the quantum-state preparation, inaccuracies in the orientation and retardation of the wave retarders, and statistical errors.

\textit{Conclusion.}
We have proposed a method for $N$-photon polarization state tomography based on entangled polarization projections. The experimental results prove the practical applicability of the proposed method with standard tools used in quantum optics laboratories. In principle, because the setup has no movable parts, all experimental error sources can be reduced to a minimum by accurately characterizing the employed optical elements. Hence, the proposed measurement strategy promises great experimental stability. The nonmechanical switching between distinct measurements can be exploited in the recently discussed Bayesian recursive data-pattern tomography \cite{Mikhalychev2015} for unprecedented speed in quantum-state reconstruction. On the practical side, this setup is well suited for miniaturization and integration in quantum optics on-a-chip experiments \cite{Sansoni2010}. Since in the case of $N=1$ the six projected states in Eq.~\eqref{eq:projectingStates} reduce to the ones used by James \emph{et al.}~\cite{James2001}, further investigations can include studies of alternative optical qubit-state tomography setups \cite{Rehacek2004} equipped with photon-number resolving detectors. Also, the question of whether quantum efficiencies of the detectors can be included in the reconstruction process and whether a reconstruction of states with indeterminate photon number is possible remains open for future related research.

\begin{acknowledgments}
This work was financially supported by the Swedish Research Council through Contract No. 621-2014-5410 and through the Linn\ae us Center of Excellence ADOPT.
\end{acknowledgments}

%

\end{document}